\documentclass[page-classic]{epl2} 
\title{Is a non-vanishing cosmological constant really needed?}
\shorttitle{Cosmologycal Constant} 

\author{J.A.Gonzalo\inst{1,2}} 
\shortauthor{J.A.Gonzalo \etal}

\institute{
  \inst{1} Divisi\'on de Ingenier\'ia, Escuela Polit\'ecnica Superior,
  Universidad San Pablo CEU, Campus Monteprincipe, Boadilla del Monte, 28668 Madrid, Spain\\
  \inst{2} Departamento de F\'isica de Materiales, C-4, Universidad Aut\'onoma de Madrid 28049 Madrid, Spain
}

\abstract{
Using consistenly the compact Friedmann-Lemaitre solutions for an open $(k=-0.580)$ universe with zero
cosmological constant ($\Lambda = 0$) we show that the oberved apparent accelerated expansion is compatible
with a time dependent Hubble parameter $H=\dot{R}/R$ equal to $H=87.95$ km/s/Mpc at $t=10.39$ Gys, $H_{0}=67.9 \pm 4
$ km/s/Mpc at $t_{o}=13.7 \pm 0.2$ Gyrs (present), $H=61.4$ km/s/Mpc at $t=15$ Gyrs and with an apparent global
flatness given by $1=\Omega_{m}+\Omega_{r}+\Omega_{k}$ at all times which, properly averaged from galaxy formation
times to present, gives $<\Omega_{m}>=0.272$, $<\Omega_{r}>=0.005$, $<\Omega_{k}>=0.728$. Therefore, no cosmological 
constant is needed.}

\begin{document}

\maketitle

\section{Introduction}
The apparent galactic accelerated expansion \cite{Perlmutter} (hinted by the observation that $\dot{r}/r$ is larger for
galaxies in our close neighborhood, $z=0.01$, than for very distant galaxies $z>1$), and the apparent global flatness
of the universe ($\Omega \simeq 1$), is currently explained in terms of "dark mass" (some $30\% $ of elusive, nonbaryonic
matter) and "dark energy" (some $70\%$ elusive energy density, somehow related to a non-vanishing cosmologycal
constant.) An intensive search of many years for direct observational evidence from dark matter and dark energy has been up to now
unsuccessful. 

On the other hand, the fact that the cosmic obervational evidence \cite{Schwarzschild,SchwarzschildII} arriving now to 
our planet from galaxies located at very different distances , i.e. from nearby galaxies ($z=0.01$) to very distant
galaxies ($z\simeq10$), necessarily involves the cosmic evolution from very early times (billions of years ago) to 
relatively recent times (only a few million years ago). This fact, in order to describe properly cosmic
dynamics, makes mandatory to take into consideration the time evolution of fundamental cosmic parameters such as the 
density parameter $\Omega(t)=\rho(t)/\rho_{c}(t)$ (where $\rho(t)$ is the matter mass density plus the radiation mass 
density and $\rho_{c}(t)=3H^{2}/8\pi G$ the critical mass density) and the product of the Hubble parameter $H=\dot{R}/R$
and $t$, the time elapsed since the big-bang, a dimensionless product which is also certainly time dependent. In most previous work this fact has not been taken properly into consideration.

We show in this work that, using consistenly the compact Friedmann-Lemaitre solutions \cite{Cereceda,Gonzalo} of Eintein's
equations for an open ($k < 0$) universe with zero cosmological constant ($\Lambda = 0$), describes very well the 
apparent accelerated expansion as well as the apparent global flatness of the universe

$1=\Omega_{m} + \Omega_{r} + \Omega_{k}$

with $<\Omega_{m}> \simeq 0.272$, $<\Omega_{r}> \simeq 0.005$, $<\Omega_{k}> \simeq 0.728$, properly averaged from the
begining of the galaxy formation time to present. 

We conclude therefore that no cosmologycal constant is needed. This amounts to a radical reinterpretation of the so called 
cosmic "dark mass" and to the so called cosmic "dark energy".

\section{A close examination to the apparent accelerated expansion}
It is well known \cite{Perlmutter} that reports of the abserved magnitude vs. redshift ($z$) for supernovae in the range 
$0.1 \leq z \leq 1$ seem to indicate a systematic cosmic acceleration.
Fig.1 depicts the expanding cosmic sphere at 

(a) $t<t_{o}$, $R<R_{o}$ : $0 \leq r \leq R$, $0 \leq \dot{r}/c \leq 1$

(b) $t=t_{o}$, $R=R_{o}$ : $0 \leq r \leq R_{o}$, $0 \leq \dot{r}/c \leq 1$

(c) $t>t_{o}$, $R>R_{o}$ : $0 \leq r \leq R$, $0 \leq \dot{r}/c \leq 1$

At present $t_{o}=13.7$ Gyrs as reported by WMAP \cite{WMAP} and anticipated by N. Cereceda, G. Lifante and J. A. Gonzalo
some years before \cite{Cereceda}, on the basis of an analysis of the calculated relationship between the dimensionless
cosmic parameters $[\Omega (y)]$ and $[H(y)t(y)]$.

The distance $r$ ($0 < r < r_{CBR}$) from the origin of the expanding background radiation sphere (our Milky Way moves
relatively very slow with respect to this origin) to a given supernova is defined in Fig.1 for cases (a), (b) and (c).
The cosmic radius (scale factor) $R$ is also given, along with the respective distance from the origin to the CBR sphere separating
the transparent universe from the plasma universe at the corresponding time. 

Therefore, at present, $r_{0}$ is the observed distance to the supernova as it was when the light wich arrives 
\textbf{now} to us was emmitted, perhaps some billion years ago, and $\dot{r_{0}}=v_{0}$ is the velocity (deduced from $z$, 
the redshift) as it was when the light which arrives \textbf{now} to us was emmitted. $R_{0}$, on the other hand is the 
cosmic radius (now) and $\dot{R_{0}}$ the speed at which the cosmic radius grows (now).

The apparent supernova distance in parsecs r(pc), where $1pc=3.07 \times 10^{8}$ cm, can be obtained \cite{Perlmutter} from

\begin{equation}
\label{eq.1}
m=5log_{10} r(pc)+(M-5)
\end{equation}

where $M=M_{sn}=-19.3$ (the relevant absolute magnitude), taking m(aparent magnitude) from Ref.1, and 
the reduced velocity $\dot{r}/c$ from

\begin{equation}
\label{eq.2}
\frac{\dot{r}}{c}= \frac{(z+1)^{2}-1}{(z+1)^{2}+1} \leq 1
\end{equation}

Note that $H$ is defined as $H=\dot{R}/R$, and the ratio $h=\dot{r}/r$ does not necessarily concide with H. 
Only for supernovae
moving away from the Milky Way in our close neighborhood ($z<<1$) is $h \simeq H$. For very distant supernovae 
(f.i. $z \simeq 1$ or larger), because of eq.(2), necessarily $h<H$.

Table I gives the reduced apparent recession velocity $(\dot{r_{0}}/c)$ vs. the apparent distance for a few representative 
supernovae \cite{Perlmutter} in the range $0.01< z < 1$. For $z$ around $z=0.01$, we have 
$\dot{r_{0}}/r_{0} \simeq \dot{R_{0}}/R_{0}=H_{0} \simeq 67.9$ km/s/Mpc, as it should. For $z \simeq 1$, on the other hand,
$\dot{r_{0}}/r_{0}$ falls increasingly below $H_{0}$, due to the fact that $\dot{r}$ remains $\dot{r} < c$, as given,
 relativistically, by eq.(2). 

This can be interpreted wrongly as a real accelerated cosmic expansion of the cosmic radius $R$, but it is not imcompatible
with a time dependent Hubble parameter $H=\dot{R}/R$ which drecreases with time, as determined below from the compact Friedmann-
Lemaitre solutions of Einstein equation's for an open universe with zero cosmological constant. In fact

\begin{equation}
\label{eq.3}
H(y)=\frac{\dot{R}}{R}=[\dot{r}/r]_{z < < 1}=\frac{c \sqrt{|k|}}{R_{+}sinh^{2}(y) tanh^{2}(y)}
\end{equation}

where $c=3 \times 10^{10} cm/s$, $\sqrt{|k|}=0.761$, $R_{+}=4.93 \times 10^{26}$ cm and $y=sinh^{-1}(R/R_{+})^{1/2}$
resulting in $c\sqrt{|k|} / R_{+} =1.515 \times 10^{3} km/s/Mpc$ if we assume that $\dot{R_{+}}=c$. 

Fig. 2 displays the data in Table I and shows that for any z, no matter how large $(\dot{r_{0}}/c)$ is less than one.
The calculated time dependence of the Hubble parameter for $t=10.3$ Gyrs, ($H=87.5$ km/s/Mpc), $t=13.7$ Gyrs, ($H=67.9$ km/s/Mpc),
$t=15.0$ Gyrs, ($H=61.4$ km/s/Mpc) indicates clearly that the rate of growth of the cosmic radius $R$ is deccelerating, which
is not incompatible with an apparent acceleration of $r$ if we come from very distant galaxies to nearby galaxies. 
$r_{o} \pm \Delta r$ for $z \simeq 0.01$ is estimated for $t < t_{0}$ and $t > t_{0}$ using

\begin{equation}
\label{eq.4}
r_{0} \pm \Delta r \simeq r_{0} \mp (\frac{\dot{r}}{c}) c \Delta t
\end{equation} 

with $\Delta t=t_{0}-t$ for $t < t_{0}$ and $\Delta t= t - t_{0}$ for $t > t_{0}$.

In particular, the dimensionless product $(Ht)$ goes up from $Ht \simeq 2/3$ ($\Omega \simeq 1$) at $t \simeq t_{CBR} \simeq
t_{af}$ (atom formation) to $H_{0}t_{0} \simeq 0.91$ ($\Omega \simeq 0.082$) at $t=t_{0}$ (now), to $Ht \simeq 1$ ($\Omega \simeq 0$)
in the relatively distant future.

According to the BLAST collaboration \cite{Schwarzschild}, $T \simeq 30K$ is a typical CBR temperature at $z > 10$ (only 
about one Gyrs after the big-bang), well below $T_{+} \simeq 60 K$, corresponding to $R_{+}=2GM/c^{2}$, 
where $M$ is the mass of the universe and $G$ is Newton's gravitational constnat, at which large
clusters of cosmic matter (then Hydrogen $H$ and primordial Helium $^{4}He$) begin to become gravitationally bound for the first time.  
At $T_{+} \simeq 60$ K ($z \simeq 21$, $t \simeq 0.365$ Gyr) the seeds of the first protostars and protogalaxies begin to grow, and,
apparently, at $T \simeq 30$ K ($z \simeq 11$, $t \simeq 0.775$ Gyrs) startbust galaxies, the so called primeval star 
nurseries, are already formed.

\section{Compact Friedmann-Lemaitre solutions}

For $k<0$, $\Lambda = 0$, Einstein's cosmological equation \cite{Cereceda} can be rewritten as

\begin{equation}
\label{eq.5}
\dot{R}=R^{-1/2}[(8 \pi /3) G \rho R^{3} + c^{2} |k| R]^{1/2}
\end{equation}

where $ \rho = \rho_{m} + \rho_{r} $ is the sum of the matter mass density and the 
radiation mass density, $c$ is the velocity of light in vacuum and $|k|$ is the absolute value of the space time curvature
(dimensionless).

We define the cosmic radius $R=R_{+}$, and the corresponding time $t=t_{+}$, as that at which

\begin{equation}
\label{eq.6}
(8 \pi /3) G \rho_{+} R_{+}^{3} = c^{2} |k| R_{+}
\end{equation}

At $R<R_{+}$ ($t<t_{+}$) the left hand term within square brakets in (5) becomes larger, and at $R>R_{+}$ ($t>t_{+}$) the
opposite is true. Then, at $t<<t_{+}$ the right hand term can be neglected, i.e. the spacetime curvature term can be neglected.
And at $t>>t_{+}$ the left hand term can be neglected and the spacetime curvature term becomes dominant. Because of that,
it is pertinent to look for solutions of eq. (5), a relatively uncomplicated nonlinear differential equation, in terms of 
$R_{+}$.

We can write the integral of eq.(5) as

\begin{equation}
\label{eq.7}
\int dt= \int \frac{R^{1/2}}{((8 \pi /3) G \rho_{+} R_{+}^{3} + c^{2} |k| R_{+})^{1/2}} dR
\end{equation}

taking into account that

\begin{equation}
\label{eq.8}
(8 \pi /3) G \rho R^{3} = 2GM =constant =c^{2}|k|R_{+} \equiv a^{2}
\end{equation}

Using the change of variable

\begin{equation}
\label{eq.9}
x^{2}=c^{2}|k|R
\end{equation}

the right hand side of eq.(7) becomes

\begin{equation}
\label{eq.10}
\int \frac{x^{2}}{a^{2}+x^{2}}dx
\end{equation}

which can be found in tables, resulting in

\begin{equation}
\label{eq.11}
t=\frac{R_{+}}{c|k|^{1/2}}=\frac{R_{+}}{c|k|^{1/2}}[sinh(y)cosh(y)-y]
\end{equation}

where $y\equiv sinh^{-1}(R/R_{+})^{1/2}$, which implies

\begin{equation}
\label{eq.12}
R=R_{+}sinh^{2}y
\end{equation}

From the compact parametric solutions giving $t(y)$ and $R(y)$ it is straightforward to get
$\dot{R}(y)$, $\rho(y)$, $\rho_{c}(y)$ and the other relevant cosmic parameters. In particular

\begin{equation}
\label{eq.13}
Ht=\frac{[sinh(y)cosh(y)-y]cosh(y)}{sinh^{3}(y)}, [\frac{2}{3} \leq Ht \leq 1] 
\end{equation}

and

\begin{equation}
\label{eq.14}
\Omega = \Omega_{m} + \Omega_{r} = \frac{1}{cosh^{2}(y)} = 1-tanh^{2}(y), [1 \leq \Omega \leq 0]
\end{equation}

For $y \rightarrow 0$ $(t\rightarrow 0)$, $Ht\rightarrow \frac{2}{3}$ and $\Omega \rightarrow 1$, while for
$y \rightarrow \infty$ $(t\rightarrow \infty )$ $Ht \rightarrow 1$, $\Omega \rightarrow 0$. 
For $y=y_{+}=sinh^{-1}(1)=0.8813$, $H_{+}t_{+}=0.753$, $\Omega_{+}=0.5$.
For $y=y_{0}=sinh^{-1}(60/2.726)^{1/2}=2.250$, $H_{0}t_{0} \simeq 0.946$, $\Omega_{0} \simeq 0.0434$.

It is clear that from early times to present times, the Hubble parameter $(H)$ and the density parameter 
$(\Omega=\Omega_{m}+\Omega_{r})$ evolve markedly. This must be taken into consideration to interpret global cosmic 
data coming from all distances and emitted at all times. 

\section{Reconsideration of the "dark matter" and "dark energy" problems taking into consideration cosmic evolution}

When we consider $\Omega= \rho_{0}/ \rho_{c0}$, what do we mean by that?, the ratio of $\rho_{0}=\rho_{m0} + \rho_{r0}$
for galaxies in our close neighborhood, or from all observable galaxies, including very distant galaxies? 
If so, allowance must be made of the fact that light from these very distant galaxies ($z \simeq 10$ or more) was emitted 
about $13 Gyrs$ ago. 

Modern treaties on cosmology postutale \cite{Weinberg}

$\Omega_{m} + \Omega_{r} + \Omega_{k} + \Omega_{\lambda}=1 $

with $\Omega_{m} \simeq 0.3$, (matter), $\Omega_{r} \simeq 0$ (radiation), $\Omega_{k} \simeq 0$ (space time curvature)
and $\Omega_{\lambda} \simeq 0,7$ (vacuum) as "realistic" values.

Let us consider quantitatively Einstein's equations using the compact Friedman-Lemaitre solutions given by eqs. (11)-(14) 
and the accurate cosmic data provided by the COBE, WMAP, HST (Hubble Space Telescope) and the reported Type Ia Supernovae 
observations, in order to get "representative" actual numbers for the various components of the density parameter $\Omega$.

Einstein's equation with $k < 0$, $ \Lambda=0 $, is given by

\begin{equation}
\label{eq.15}
\dot{R}^{2}=\frac{8 \pi G}{3}\rho R^{2} + c^{2}|k|
\end{equation}

which is the same as eq.(5), obtained eliminating $\dot{R}$ and rearanging terms.

Eq.(15) is obviosly equivalent to

\begin{equation}
\label{eq.16}
\frac{3(\dot{R}/R)^{2}}{8 \pi G}=\rho + \frac{3(\dot{R}/R)^{2}}{8 \pi G} |k| (\frac{c}{\dot{R}})^{2}
\end{equation}

and, dividing both sides by $\rho_{c}=3(\dot{R}/R)^{2}/8 \pi G$, 

\begin{equation}
\label{eq.17}
1=(\Omega_{m} + \Omega_{r})+ \Omega_{k}
\end{equation}

where $\Omega_{k} \equiv |k|(\frac{c}{\dot{R}})^{2}$ has been used.

Tables II and III give the time evolution of the following quantities

\begin{equation}
\label{eq.18}
\Omega_{m} = \Omega/(1+ \frac{T}{T_{af}})
\end{equation}

\begin{equation}
\label{eq.19}
\Omega_{r} = \Omega (\frac{T}{T_{af}})/(1+ \frac{T}{T_{af}})
\end{equation}

\begin{equation}
\label{eq.20}
\Omega = \frac{1}{cosh^{2}y}= 1-tanh^{2}y
\end{equation}

\begin{equation}
\label{eq.21}
\Omega_{k} = 1- \frac{1}{cosh^{2}y}= tanh^{2}y
\end{equation}

\begin{equation}
\label{eq.22}
z=\frac{T}{T_{0}}-1
\end{equation}

\begin{equation}
\label{eq.23}
y=sinh^{-1}(T_{+}/T)^{1/2}=sinh^{-1}(R/R_{+})^{1/2}
\end{equation}

\begin{equation}
\label{eq.24}
t=\frac{R_{+}}{c|k|^{1/2}}[sinh(y)cosh(y)-y]
\end{equation}

where $T$ is the CBR temperature at time $t$, $T_{af}=2968$ K (temperature at atom formation), $T_{+}=59.2$ K
(temperature at which $\Omega_{m}+\Omega_{r}=\Omega_{k}=0.5$), $T_{0}=2.726$ K (COBE), $R_{+}=4.93 \times 10^{26}$ cm
and $|k|^{1/2}=0.761$, obtained using \cite{WMAP} $H_{0}=67$ km/s/Mpc and $t_{0}=13.7$ Gyrs which imply 
$\Omega_{0}=\Omega_{m0}+\Omega_{r0}=0.044$, with estimated errors for these quantities of the order of $6 \%$ or less.

In the present transparent \cite{Cereceda}, matter dominated phase of the cosmic expansion ($RT=constant$) the ratio of 
$\rho_{r}$ to $\rho_{m}$ is given by

\begin{equation}
\label{eq.25}
\frac{\rho_{r}(T)}{\rho_{m}(T)}=\frac{\rho_{raf}(T/T_{af})^{4}}{\rho_{maf}(T/T_{af})^{3}}=\frac{T}{T_{af}}
\end{equation}

after taking into account that $\rho_{raf} \simeq \rho_{maf}$ at decoupling (equality, atom formation). 

$\Omega_{m}$ decreases with time from $\Omega_{m+} \simeq 0.490$ at the time when the first galaxies begin to form
($t_{+} \simeq 0.365 Myrs$) to thepresent time ($t_{0} = 13.7 Gyrs$) at which it becomes $\Omega_{m0} \simeq 0.044$.
$\Omega_{r}$ in this time interval is a small fraction of $\Omega_{m}$ all the way. At an 
earlier time ($t_{af} \simeq 1.28$ Myrs) however, i.e. at atom formation time, $\rho_{r}$ becomes equal to $\rho_{m}$
and $\Omega_{r} \simeq \Omega_{m} \simeq 1/2$. $\Omega_{k}$ on the other hand increases from $\Omega_{k+} \simeq 0.500$ at
$t_{+}$ to $\Omega_{k0} \simeq 0.965$ at $t_{0}$, and according to Table III was very small $\Omega_{kaf} \simeq 0.019$,
at $t_{af}$ (atom formation).

The time evolution of $\Omega_{k}(t)$ is therefore opposite to that of $\Omega (t) = \Omega_{m} (t)+ \Omega_{r} (t)$
(see Fig.3). 

"Representative" (average) numbers from $t=t_{+}=0.365$ Gyrs to the present time $t=t_{0}=13.7$ Gyrs are the following

\begin{equation}
\label{eq.26}
<\Omega_{m}> \simeq 0.272
\end{equation} 

\begin{equation}
\label{eq.27}
<\Omega_{r}> \simeq 0.005
\end{equation} 

\begin{equation}
\label{eq.28}
<\Omega_{k}> \simeq 0.728
\end{equation} 

This implies $<\Omega_{\Lambda}>=0$ for the "vacuum" contribution to the density parameter. It gives also a completely
different new meaning to the "dark matter" and "dark energy" contributions.

In view of these considerations it may be concluded that, after all, Einstein was not far from the truth \cite{Bazeia}
when he said that by introducing the cosmological constant he made the greatest blunder in his scientific career.

In writting the final version of this paper we became aware of recent work \cite{Bazeia} on dark energy and dust for
cosmological models described by a real scalar field in the presence of dust in spatially flat space. It may be noted that
in Fig.1 of Ref.\cite{Bazeia} $\Omega_{\phi}$ and $\Omega_{d}$ given as a function of the scale factor $a$ resemble closely
$\Omega_{k}$ and $\Omega=\Omega_{m}+\Omega_{r}$ as given in this work as a function of $y \equiv sinh^{-1}(R/R_{+})^{1/2}$.
It may be noted also that in the framework of our compact Friedmann-Lemaitre solutions the decelerating parameter $q$ is
equal to $q \equiv \ddot{R} R/ \dot{R} ^{2} \simeq \frac{1}{2}(1-tanh^{2}(y))$ as already noted in previous work \cite{GonzaloII}.

\begin{figure}
\includegraphics[width=12cm,height=12cm,angle=-90]{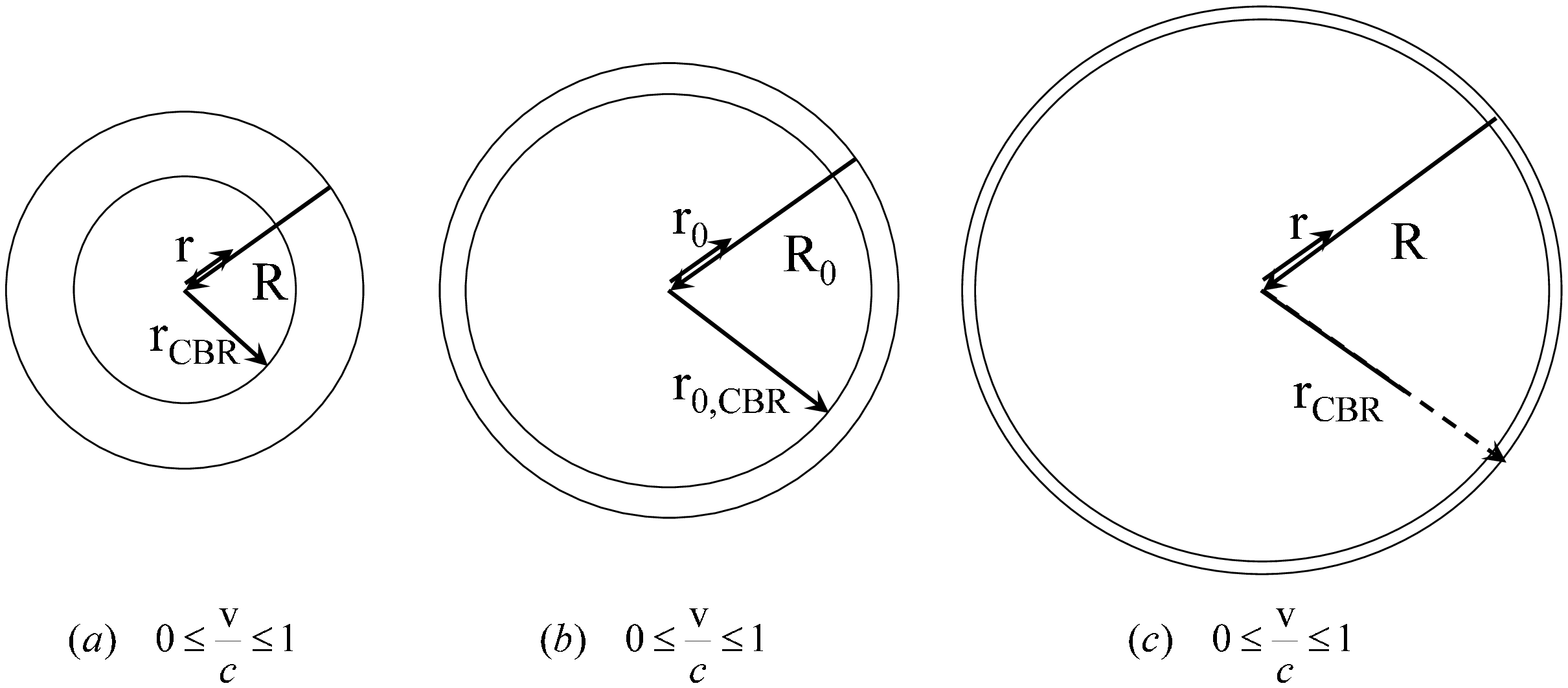}
\caption{Expading cosmic sphere at (a) $t<t_{0}$; (b) $t=t_{0}$; (c) $t>t_{0}$
}
\label{fig1}
\end{figure}

\begin{figure}
\includegraphics[width=12cm,height=12cm,angle=-90]{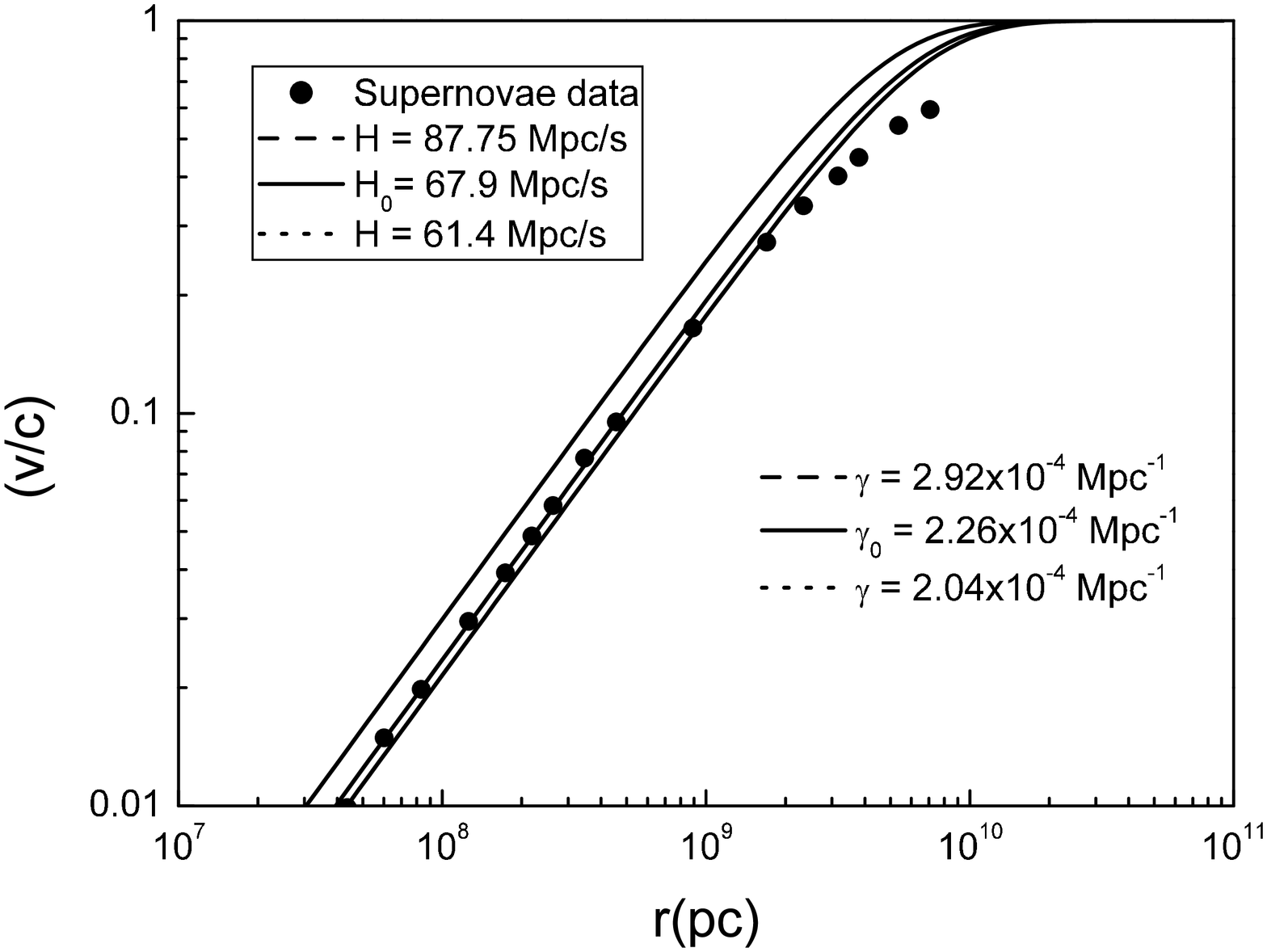}
\caption{Apparent velocity ($\dot{r}/c$) vs. apparent distance ($r(pc)$) fitted by means of $v/c \simeq tanh(\gamma r)$.
}
\label{fig2}
\end{figure}

\begin{figure}
\includegraphics[width=12cm,height=12cm,angle=-90]{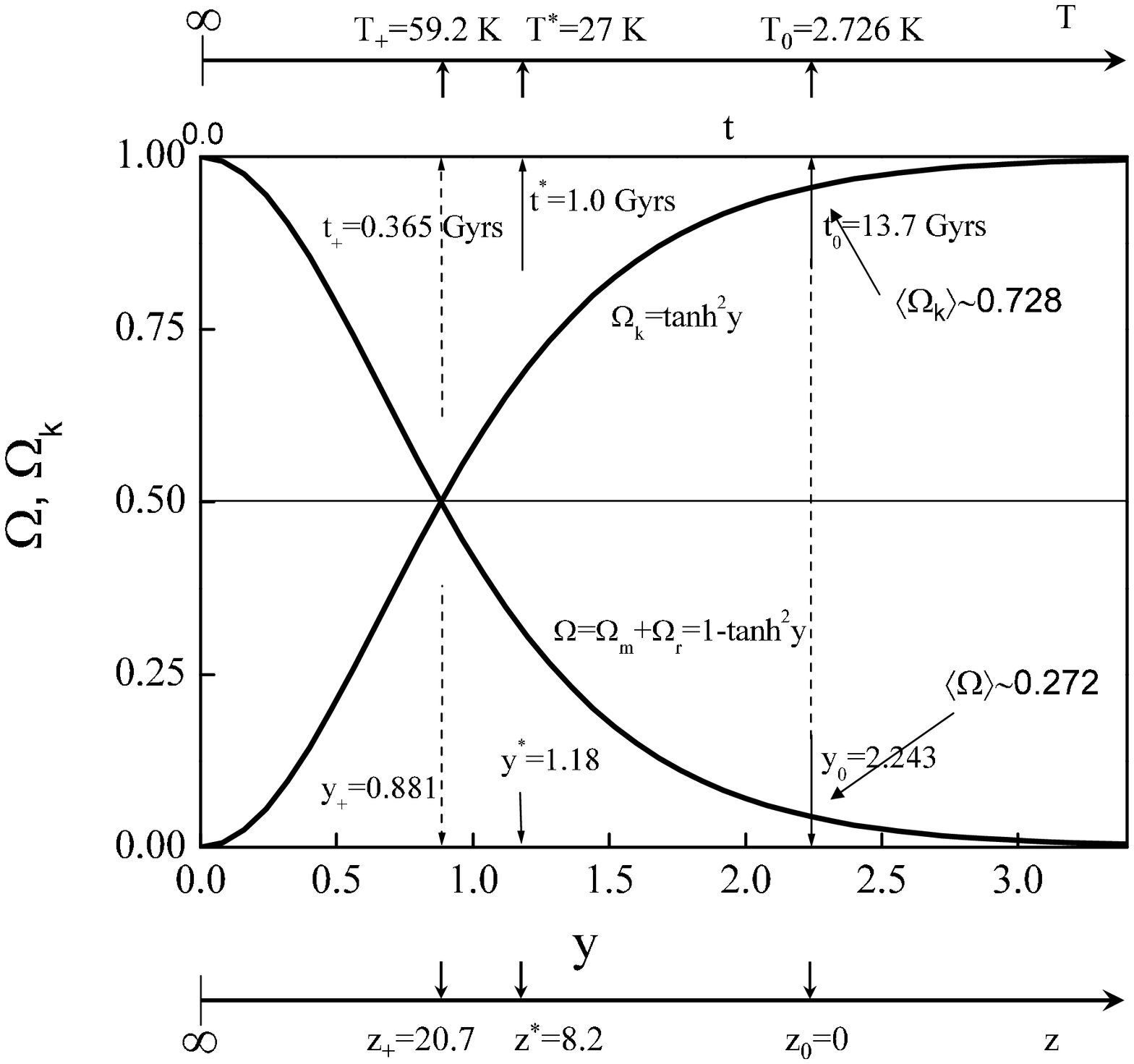}
\caption{Contributions to the density parameter $\Omega_{m}$ (matter), $\Omega_{r}$ (radiation), $\Omega_{k}$ (spacetime
curvature) as a function of $y=sinh^{-1}(T_{+}/T)^{1/2}=sinh^{-1}(R/R_{+})^{1/2}$, $z$ (redshift), $t$ (time)
and $T$ (CBR, Cosmic Background Radiation temperature).
}
\label{fig3}
\end{figure}

\begin{table}
\caption{Apparent velocity ($\frac{\dot{r}_{0}}{c}$) and apparent distance $r_{0} (pc)$ for supernovae
with $z$ in the range $0.01<z<1$ }
\label{tab.1}
\begin{center}
\begin{tabular}{lcr}
Redshift $z$  & $\frac{\dot{r}_{0}}{c}=\frac{(z+1)^{2}-1}{(z+1)^{2}+1}$ & $r_{0}(pcs) \times 10^{7}$ \\
0.010 & 0.0099 & 4.36\\
0.015 & 0.0149  & 6.02\\
0.020 & 0.0198 & 8.31\\
0.030 & 0.0295 & 12.58\\
0.040 & 0.0392 & 17.37\\
0.050 & 0.0487 & 21.87\\
0.060 & 0.0582 & 26.30 \\
0.080 & 0.0768 & 34.67\\
0.100 & 0.0950 & 45.70\\
0.181 & 0.1648 & 89.10\\
0.323 & 0.2728 & 169.80\\
0.421 & 0.3375 & 234.40 \\
0.532 & 0.4024 & 316.20 \\
0.620 & 0.4481 & 380.00\\
0.835 & 0.5408 & 549.00 \\
0.980 & 0.5935 & 724.00\\
\end{tabular}
\end{center}
\end{table}

\begin{table}
\caption{Cosmological parameters $t$ (time) and $T$(CBR temperature) as a funtion of $y$ ($z=$redshift)}
\label{tab.2}
\begin{center}
\begin{tabular}{lcr}
y(z) & t(Gyrs) & T(K)\\
2.243(0) & 13.70 & 2.726 \\
2.196(0.100) & 12.36 & 3.000 \\
2.057(0.467) & 9.08 & 4.000 \\
1.949(0.834) & 7.14 & 5.000 \\
1.862(1.201) & 5.83 & 6.000 \\
1.789(1.567) & 4.92 & 7.000 \\
1.726(1.934) & 4.23 & 8.000 \\
1.671(2.301) & 3.70 & 9.000 \\
1.400(4.990) & 1.85 & 16.33 \\
1.200(8.53)  & 1.05 & 26.00 \\
1.100(11.18) & 0.775 & 33.22 \\
\bf{0.881(20.72)} & \bf{0.365} & \bf{59.23} \\
0.800(26.54) & 0.266 & 75.09 \\
0.700(36.75) & 0.173 & 102.92 \\
0.600(52.60) & 0.106 & 146.20 \\
0.400(127.80) & 0.030 & 351.23 \\
-------- & --- & --- \\
0.140(1087.0) & 0.001 & 2968.00 \\
\end{tabular}
\end{center}
\end{table}

\begin{table}
\caption{Contributions to the density parameter $\Omega_{m}$ (matter), $\Omega_{r}$ (radiation), $\Omega_{k}$ (space-time curvature)
as a funtion of $y=sinh^{-1}(T_{+}/T)$}
\label{tab.3}
\begin{center}
\begin{tabular}{lcr}
$\Omega_{m}$ & $\Omega_{r}$ & $\Omega_{k}$ \\
0.044 & --- & 0.956 \\
0.048 & --- & 0.951 \\
0.063 & --- & 0.936 \\
0.077 & --- & 0.922 \\
0.091 & --- & 0.908 \\
0.105 & --- & 0.895 \\
0.118 & --- & 0.881 \\
0.131 & --- & 0.869 \\
0.215 & 0.001 & 0.783 \\
0.302 & 0.002 & 0.695 \\
0.355 & 0.004 & 0.640 \\
\bf{0.490} & \bf{0.009} & \bf{0.500} \\
0.545 & 0.013 & 0.440 \\
0.613 & 0.021 & 0.365 \\
0.678 & 0.033 & 0.288 \\
0.765 & 0.090 & 0.144 \\
----- & ----- & ----- \\
0.490 & 0.490 & 0.019 \\
\end{tabular}
\end{center}
\end{table}

\acknowledgments
We (specially JAG) are grateful to Gines Lifante, Manuel I. Marqu\'es, Manuel M. Carreira, Stanley L. Jaki, Ralph A. Alpher, John C. Mather, Dermott Mullan
and Anthony Hewish, among others, for helpful comments and encouraging correspondence through the years.

\end{document}